\documentstyle[12pt]{article}
\renewcommand{\textwidth}{16cm}
  \newlength{\absize}
\begin{document}
\newcommand\Bkg{$B \rightarrow K^\ast \; \gamma\,$ decay}
\newcommand\Brg{$B \rightarrow \rho \; \gamma\, $ decay}
\newcommand\bsg{$b \rightarrow s \; \gamma\,$ decay}
\newcommand\bdg{$b \rightarrow d \; \gamma \,$ decay}
\newcommand\Bdg{$B \rightarrow  X_d \; \gamma \,$ decay}
\newcommand\gev{{\rm GeV }}
\newcommand\mev{{\rm MeV }}
\newcommand\vcb{$V_{cb} $}
\newcommand\vtb{$V_{tb} $}
\newcommand\vud{$V_{ud}^\ast $}
\newcommand\vcd{$V_{cd}^\ast $}
\newcommand\vtd{$V_{td}^\ast $}
\newcommand{\preprint}[1]{%
  \begin{flushright}
   \setlength{\baselineskip}{3ex} #1
  \end{flushright}}
\renewcommand{\title}[1]{%
  \begin{center}
    \LARGE #1
  \end{center}\par}
\renewcommand{\author}[1]{%
  \vspace{2ex}
  {\normalsize
   \begin{center}
     \setlength{\baselineskip}{3ex} #1 \par
   \end{center}}}
\renewcommand{\thanks}[1]{\footnote{#1}}
\renewcommand{\abstract}[1]{%
  \vspace{2ex}
  \normalsize
  \begin{center}
    \centerline{\bf Abstract}\par
    \vspace{2ex}
    \parbox{\absize}{#1\setlength{\baselineskip}{2.5ex}\par}
  \end{center}}
\begin{titlepage}
\preprint{HUTP-94/A037\\DSF-T-95/2\\hep-ph 9502286}
\vfill
\title{
Short and Long Distance
Interplay \\ in Inclusive  $ B \rightarrow  X_d \gamma$
 Decay}
\vskip 1.6cm
\author{{\large Giulia Ricciardi}\\\hfil\\
Lyman Laboratory of Physics~\thanks{ e-mail address:
{\tt ricciardi@physics.harvard.edu}}\\
        Harvard University \\ Cambridge, MA 02138\\
and  \\
        Dipartimento di Scienze Fisiche\\
        Universit\`a  degli Studi di Napoli\\
        Mostra d' Oltremare, Pad 19\\
       I-80125 Napoli.}
\vfill
\abstract{We analyze
the short and long
distance contributions to
inclusive $B \rightarrow X_d \gamma$ decay,
paying particular attention to the dependence on the
Cabibbo-Kobayashi-Maskawa parameter $V_{td}$.
We discuss  penguin diagrams with internal
$u$ and $c$ quarks in the framework of the
effective field theory.
We also estimate the size of possible long range contributions
by using vector meson dominance.}
\vskip2truecm
\centerline{\sl To be published in Phys. Lett. B}
\vfill
\end{titlepage}
\vskip2truecm
\section{Introduction}
\label{sec:intro}

The radiative penguin decays
 have
 recently received considerable theoretical attention.
 The short distance QCD corrections to
$b \rightarrow s \,\gamma$  decay have been
calculated completely at the leading order~\cite{bsg:LO}
and partially at the next-to-leading
order~\cite{ ali-greub:ntl};
the branching ratio is in agreement
(within the errors)
with the experimental
value from
CLEO~\cite{CLEO}.
 The penguin diagrams
for the $b \rightarrow s \,\gamma$
decay are dominated by the virtual $t$-quark contribution
that is proportional to the element of the
Cabibbo-Kobayashi-Maskawa (CKM) matrix  $V_{ts}$.
On the other hand, the $b \rightarrow d \,\gamma$
decay can, if dominated by the
$t$-quark loop,  provide
information on $V_{td}$.
This
decay has not been detected
yet. The expected branching ratio
is approximately $10^{-5}$
(see e. g.  \cite{ali-greub:inclusive})
and the statistics required is
 of order $10^{7}-10^{8}$ $B$ mesons,
which is within the reach of future
CLEO/B factories.
The principal
experimental challenge is to keep
 a  good control of the
main
background due to the \bsg.

In this paper we will analyze the inclusive
$B \rightarrow X_d\, \gamma$ decay.
It has been suggested (see e.g.~\cite{soni-vtd}) that
its amplitude
may not proportional to $|V_{td}|$
because of sizable contributions
from penguin diagrams with internal $c$
and $u$ quarks.
We compare
the relative contributions
of penguin diagrams
with heavy and light loops
in the framework of the effective
field theory.
We are
interested whether the dependence
 on the light
masses is logarithmic or powerlike,
since we expect that terms with
powerlike dependence
are negligible.
We also aim to
estimate possible long range contributions to the
decay rate of
inclusive $B \rightarrow X_d \, \gamma$;
we adopt the phenomenological approach
of vector meson dominance (VMD).

\nopagebreak
\section{The Short Distance}
\label{short-distance}

The amplitude $A$ for the
$b \rightarrow s \; \gamma \,$ or $b \rightarrow d \; \gamma \,$
 decay can be written as
$
A~=~v_t~A_t~+~v_u~A_u~+~v_c~A_c,
$
where
$v_t = V_{tb}\, V_{tx}^\ast,\,
v_c = V_{cb}\, V_{cx}^\ast,\,
v_u =V_{ub}\, V_{ux}^\ast$, with $x=s$ or $d$ respectively.
By using the unitarity of the CKM matrix,
we can rewrite the amplitude in the form
\begin{equation}
A = v_t (A_t -A_c) + v_u (A_u - A_c).
\label{amplitude}
\end{equation}

In the case of $b \rightarrow s \; \gamma \,$,
 we have $v_t \sim O(\lambda^2)$
and $v_u \sim O(\lambda^5)$,
where  $\lambda \simeq 0.22$
is the Cabibbo suppression factor;
 thus, $v_u \ll  v_t $ and it is
generally considered safe to ignore the
second term in (\ref{amplitude}).
For the $b \rightarrow d \; \gamma \,$,
however, $v_t$ and $v_u$ are comparable
$\sim O(\lambda^3)$;
that prompt us to examine more carefully
the size of $A_u-A_c$.

Let us first consider the $b \rightarrow d \; \gamma \,$
amplitude in the absence of QCD corrections.
The gauge invariant form for $A_{u,c,t}$
resulting from the expansion
to the second order in the external momenta is
\begin{equation}
A_q \equiv A_q^P+A_q^M  \qquad \qquad
q =\{ u, c,t \},
\label{sd-am-ap1}
\end{equation}
with
\begin{eqnarray}
A^P_q &\equiv&
F_P(m^2_q/m^2_W) \;\bar d (\hat{k} k^\mu -k^2 \gamma^\mu)
\, (1-\gamma_5) b
\, \epsilon^\gamma_\mu
\nonumber \\
A^M_q &\equiv& F_M(m^2_q/m^2_W)
\; \bar d \sigma^{\mu\nu} k_\nu [m_d (1-\gamma_5) + m_b
(1+\gamma_5)] b
\, \epsilon^\gamma_\mu,
\label{sd-am-ap}
\end{eqnarray}
where
$\epsilon^\gamma_\mu$
is the photon polarization vector,
$k$ is the photon momentum
and  the coefficients
$F_P, \, F_M$ are independent of the momenta.
$A_q^P$ and $A_q^M$ will be referred
as the penguin and magnetic moment amplitudes.
As worked out explicitly in
\cite{InamiLim},
the
 dependence of $A^M_q$ on
the ratio $m_q^2/m_W^2$
is powerlike, while in $A^P_q$
there are also logarithmic terms.
This can be qualitatively
understood in the following way.
By naive power counting in the euclidean space,
we observe that,
before the expansion in the external momenta,
the one loop diagrams have no IR divergence
because of the massive $W$ propagator.
After the expansion,
the only source of infrared (IR) divergence in the one loop diagrams is
a logarithmic dependence
on the internal quark masses,
when the masses go to zero.
The expansion induces
the  following IR behaviour,
expressed symbolically
in terms of the internal momentum $l$
\begin{equation}
A_q^P  \sim  k^2 \int d^2l \frac{1}{l^2} \, , \qquad \qquad
A_q^M  \sim  k_\nu \int d^3l \frac{1}{l^2}.
\label{IR12}
\end{equation}
Therefore we expect
 IR divergence (and consequently logarithms in the light masses)
only in the
penguin terms~\footnote{Note that this is not necessarily true in
diagrams with more than one loop, since the presence
of  more
internal loop momenta and of IR subdivergences alter
the naive power counting of
{}~(\ref{IR12}).}.
When the photon is real,
 only the amplitudes $A^M_q$ contribute to
the decay.
 On the other hand, $A^M_c$ and $A^M_u$
are strongly suppressed due to their powerlike dependence
on ${m_q^2}/{m_W^2}$.
By using the results of Inami and Lim~\cite{InamiLim},
we find
\begin{equation}
\frac{A_c - A_u}{A_t - A_c} =
\frac{A_c^M - A_u^M}{A_t^M - A_c^M} \simeq 5  \times 10^{-4}
\qquad
(m_t=174\; \gev\;, \;m_c=1.5 \;\gev).
\label{ratio-light-heavy}
\end{equation}
Therefore, in the absence of QCD corrections,
it is justified to assume proportionality to $v_t$  in the
 $b \rightarrow d \; \gamma \,$ decay.

In order to  include QCD corrections,
we introduce
the effective field theory formalism
and work at the lowest order in the
weak interactions.
The basis of operators
 $\{Q_1...Q_8\}$
for the $b \rightarrow s \, \gamma$
decay is well known
and we will not explicitly reproduce it here; the reader is
referred to \cite{bsg:LO}.
After the substitution
of the $s$-quark with the $d$-quark,
 all the operators in this basis become suitable
for the $b \rightarrow  d \; \gamma \,$ decay.
In addition, we have two new current-current
operators
generated
by integrating out the $W$-boson
\begin{eqnarray}
Q^u_{1} &=& {\bar d}_{\alpha} \gamma_\mu (1-\gamma_5)
u_{\beta}\;
    {\bar u}_{\beta} \gamma_\mu (1-\gamma_5) b_{\alpha}
   \nonumber\\
Q^u_{2}&=&{\bar d}_{\alpha} \gamma_\mu (1-\gamma_5) u_{\alpha}\;
    {\bar u}_{\beta} \gamma_\mu (1-\gamma_5) b_{\beta}.
  \label{basis-2}
\end{eqnarray}
Since the effective theory and
the full theory share, by definition,
the same low energy behaviour,
all possible
 logarithmic singularity in the light masses (for
$m_c$, $m_u \rightarrow 0$)  are cancelled
at the matching.
Therefore, at the matching scale $\mu=M_W$,
the coefficients
 of the effective hamiltonian
may have  a powerlike and logarithmic dependence
on the
heavy masses, but only
 a  powerlike dependence on the light masses.
The only non-zero coefficients are
\begin{eqnarray}
C_2(m_W) &=& v_c \nonumber \\
C_7(m_W) &=& v_t
\left[ \frac{1}{2} \,  F_2\left(
\frac{m^2_t}{m^2_W}\right)-
O\left(\frac{m^2_c}{m^2_W}\right) \right]+
v_u \left[O \left(\frac{m^2_u}{m^2_W}\right) -
O\left(\frac{m^2_c}{m^2_W}\right)\right]
\nonumber \\
&\rightarrow&
v_t \frac{1}{2}
\, F_2\left(\frac{m^2_t}{m^2_W}\right)
\nonumber \\
C_8(m_W) &=& v_t  \left[
\frac{1}{2}\,\tilde F_2\left(\frac{m^2_t}{m^2_W}\right)-
O\left(\frac{m^2_c}{m^2_W}\right) \right]+
 v_u \left[O \left(\frac{m^2_u}{m^2_W}\right) -
O\left(\frac{m^2_c}{m^2_W}\right)\right]
\nonumber \\
&\rightarrow&
v_t \frac{1}{2}
\, \tilde F_2 \left(\frac{m^2_t}{m^2_W}\right)
\nonumber \\
C^u_{2}(m_W) &=& v_u
\label{sd-coefficients}
\end{eqnarray}
where $F_2$ and $\tilde F_2$
are Inami-Lim coefficients \footnote{
$F_2$ is  given by  Eq.
(B.3) in~\cite{InamiLim}.
$\tilde F_2$ comes from the diagrams
$(a)$, $(b)$, $(c)$ and $(d)$
in~\cite{InamiLim}
when the $Z$ is replaced by a gluon;
therefore $F_2 = Q \tilde F_2 + ...$.}.
The QCD  rescaling
does not  change the powerlike dependence
of the amplitude
on the light masses.
The
 anomalous dimension matrix
 for  $b \rightarrow  d \; \gamma \,$
differs from
 the anomalous dimension matrix
for
$b \rightarrow s \; \gamma \,$
 only for
 the additional
 entries due to
$Q_1^u$ and $Q_2^u$.
Thanks to their similarity with $Q_1$ and $Q_2$,
$Q_1^u$ and $Q_2^u$ mix with all the other operators
in the same way.
Due to the flavor difference,
they
do not mix with $Q_1$ and $Q_2$.
In $b \rightarrow s \; \gamma $
and $b \rightarrow d \; \gamma $,
the amplitude is proportional to
the so-called effective coefficient
$C_7^{\rm eff}$
(see e.g.~\cite{buras:theor-uncertainties}),
that includes finite contributions from matrix elements.
At the leading order
and
in naive dimensional regularization,
the only non-zero
one-loop matrix elements
are the matrix elements of
$Q_3, Q_4, Q_5, Q_6$ with a massive
internal $b$-quark.
These finite terms
have been calculated in the
$b \rightarrow s \; \gamma \,$  case and
are  obviously left invariant  by
the inclusion of $Q_1^u$ and $Q_2^u$
in the basis. There are no other possible
sources of logarithms in the
light masses at the leading order;
as a result,
$C_7^{eff}$ is
\begin{eqnarray}
C_7^{\rm eff}(\mu) &=& \eta^{16/23} C_7(m_W) + \frac{8}{3}
\left( \eta^{14/23} -\eta^{16/23} \right)
C_8(m_W) + \nonumber \\
& & (C_2(m_W) + C_{2}^u(m_W))
\sum_1^{8} h_i \eta^{a_i},
\label{vaffa}
\end{eqnarray}
where $\eta = \alpha_s(m_W)/\alpha_s(\mu)$.
The numbers $a_i$ and $h_i$ are given
by the eigenvectors
and eigenvalues of the anomalous dimension
matrix~\cite{bsg:LO}.
By equation (\ref{sd-coefficients})
and using the unitarity of the CKM matrix, we can see
that the expression is still proportional to $v_t$,
as pointed out also in~\cite{ali:sum-rules}.

It must be stressed that
this result
is peculiar to   $b \rightarrow d \; \gamma$  (and  to
$b \rightarrow s \; \gamma$). For instance,
in decays like $b \rightarrow (d, s) \, q \,\bar q$
there are non-zero  penguin diagrams
with a virtual photon and massless internal
quarks.
They have
a logarithmic dependence on the light masses and introduce
non-negligible contributions
proportional to $v_u$ in the amplitude~\cite{buras:angle-beta}.

At the next-to-leading order the situation is different.
Now  the matching  is performed at
$\alpha_s$ order with two loops diagrams.
The expansion in the external momenta of the two loops diagrams
gives rise to penguin and magnetic moment amplitudes
like in (\ref{sd-am-ap1}) and (\ref{sd-am-ap}).
Since the naive power counting in (\ref{IR12})
does not necessarily work at two loops,
both amplitudes
 may have
a logarithmic dependence on the internal masses.
We expect the singularity in the light masses to
 be
reproduced in the effective theory by the matrix elements
of the four quark operators.
This is similar to what happens in $b \rightarrow
s \, e^+\, e^-$;
the logarithms cancelled at the matching in the
coefficient of the operator $ \bar s \gamma_\mu (1-\gamma_5) b
\; \bar e \gamma^\mu e $
are recovered at lower energy in the matrix elements
of $H_{eff}$~\cite{grinstein:bsee}.
At the same way,
in $b \rightarrow d \; \gamma$,
nothing prevents the  two loop matrix elements of
$Q_1, Q_2, Q_1^u, Q_2^u$
(proportional to $v_c$ and $v_u$)
to have a non-negligible  logarithmic dependence on the light masses.
Unfortunately,
the next-to-leading calculation has not been
completed  even for
the \bsg,
making  a reliable  estimate
difficult;
roughly speaking,
we expect
non-negligible
terms proportional to $v_u$
in~(\ref{amplitude}) to be
suppressed by an extra power
of $\alpha_s$.
Since $\alpha_s(m_b) \sim 0.21$,
however, this contribution may well be sizable.

In the inclusive $B \rightarrow X_d \, \gamma $ decay
one has to take into account
also the gluon
bremsstrahlung corrections via the
$b \rightarrow d + g +\gamma$
decay.
These corrections have been
partially calculated in
\cite{ali-greub:inclusive}.
The matrix elements of the
four fermion operators in
$b \rightarrow d + g +\gamma$ introduce  additional
terms that are not proportional to $V_{td}$
with a
logarithmic dependence
on the light masses.

\nopagebreak
\section{The Long Distance}
\label{long-distance}

The short distance approach leaves
out possible contributions
due to intermediate
hadronic states.
These contributions
may be estimated
by using the vector meson dominance (VMD)
hypothesis,
as suggested by~\cite{Deshpande:LDPenguins}.
According to VMD, the
 $b \rightarrow d \, \gamma$
 decay is described
by the  $b \rightarrow d \, V$ decay, where $V$
is a vector meson,
($\psi$ and its excited states,
$\rho$ and $\omega$),
followed by
the conversion $V \rightarrow \gamma $.
Since the applicability itself of VMD  is not
well established in the case of radiative penguin decays,
we will limit ourselves  to give
an order of magnitude of this contribution
and
we will  make very simple hypothesis.
By Lorentz invariance,
the most general interaction
proportional to
the vector meson polarization $\epsilon^V_\mu$
is given by
\begin{eqnarray}
A &\propto & \epsilon_\mu^V
\bar d \{
\gamma^\mu [ a_1 \, (1-\gamma_5) +
b_1 \, (1+\gamma_5) ]   \nonumber \\
&+&
\sigma^{\mu\nu}
[ a_2\, (1-\gamma_5) + b_2 \, (1+\gamma_5) ]
p_\nu +  \sigma^{\mu\nu}
[ a_3 \, (1-\gamma_5) + b_3 \, (1+\gamma_5) ]
 q_\nu \nonumber \\
&+&
 [ a_4 \, (1-\gamma_5) + b_4 \, (1+\gamma_5) ] p^\mu
+  [ a_5 \, (1-\gamma_5) + b_5 \, (1+\gamma_5) ] q^\mu
\} b,
\end{eqnarray}
where $ p_\mu \equiv p^d_\mu + p^b_\mu$
is the sum of $d$ and $b$ momenta and
 $ q_\mu \equiv p^d_\mu - p^b_\mu $ is the
vector meson moment.
Terms proportional to $\bar s \, q^\mu (1 \pm \gamma_5)\,b $ are zero
since the vector meson is on-shell
and
$
\epsilon_\mu^V q^\mu =0
$.
Since the
quarks are on shell, we
may use the Gordon decomposition to write
\begin{eqnarray}
A &\propto& \epsilon_\mu^V
\bar d \{
\sigma^{\mu\nu}
[ a_2\, (1-\gamma_5) + b_2 \, (1+\gamma_5) ]
p_\nu + \sigma^{\mu\nu}
[ a_3 \, (1-\gamma_5) + b_3 \, (1+\gamma_5) ]
q_\nu  \nonumber \\
&+&  [ a_4 \, (1-\gamma_5) + b_4 \, (1+\gamma_5) ] p^\mu \} b.
\label{amplitude:gauge-inavariant}
\end{eqnarray}
The  gauge invariance
requires that
 only
transverse terms
couple to
the photon;
therefore, we obtain  the  following
amplitude
\begin{equation}
A_T( b \rightarrow d \, V)
= \frac{G_F}{\sqrt{2}}
v_q
 f_V(m_V^2) \frac{m_V}{m_b}
\left[
a_3 \, \bar d {\sigma^{\mu\nu}} (1-\gamma_5) \,q_\nu  b
+ b_3 \, \bar d {\sigma^{\mu\nu}} (1+\gamma_5) \,q_\nu b \right]
 \epsilon_\mu^V ,
\label{Atransverse}
\end{equation}
where $f_V$ is
the $V$ decay constant
(see Table 1)
and $e_V$
is the charge factor of the
constituent states according to the quark model
($e_\rho = 1/\sqrt{2}$, $e_\omega = 1/(3\, \sqrt{2})$,
$ e_\psi = 2/3 $).
The two terms
in (\ref{Atransverse}) do not interfere ($m_d\simeq0$)
and give the same contribution to the transverse rate:
$ |A_T(b \rightarrow s\,\gamma)|^2 \propto
\xi_V^2$ where
$\xi_V \equiv \sqrt{|a_3|^2+|b_3|^2}$.

Using the measured transverse decay rate for
 $B \rightarrow \psi\, X$, we can fit
$\xi_\psi$.
Experimentally~\cite{stone:transverse,balest}
\begin{eqnarray}
Br (B \rightarrow \psi + X) &=& 0.81\pm 0.08 \%
\label{brpsi} \\
\Gamma_L/\Gamma &=& 0.78 \pm 0.17  \qquad ( 1.4  \,{\rm GeV}
< p_\psi < 2.0 \,{\rm GeV}) \label{long:high}, \\
\Gamma_L/\Gamma &=& 0.59 \pm 0.15
\qquad  (p_\psi < 2.0 \,{\rm GeV}),
\label{long:low}
\end{eqnarray}
where $\Gamma_L$ is the longitudinal decay rate.
The branching ratio~(\ref{brpsi})
 is
for direct production of $\psi$,
while the polarization data~(\ref{long:low}) and~(\ref{long:high})
include  decay products of heavier charmoniums.
It is known~\cite{balest} that the momentum range
$1.4 \,{\rm GeV} < p_\psi < 2.0 \,{\rm GeV}$
is dominated by the exclusive
modes $\psi \, K (K^\ast) $,
 while the low momentum region
$p_\psi < 1.1 \,{\rm GeV}$
is mostly
from heavier charmoniums.
 We expect~(\ref{long:high}) to be
 an overestimate since $\psi K$
is purely longitudinal by kinematics, and~(\ref{long:low})
 to be an underestimate since it
is diluted by decay products. Here, we simply take the average and use
$
\Gamma_L /\Gamma = 0.7
$
for direct inclusive $\psi$ production.
Using Eqs~(\ref{Atransverse}), (\ref{brpsi}) and
$\Gamma_T/\Gamma = 1-
\Gamma_L / \Gamma = 0.3$
we obtain
$
\xi_{\psi} \simeq 0.19.
$

The measured branching ratio
for the inclusive $B$ decay into $\psi^\prime$
is
$
{\rm Br( B} \rightarrow \psi^\prime +X ) = (0.34\pm 0.04
\pm 0.03 ) \, \%
$~\cite{balest}.
We assume that the measured branching ratio
is equal to the direct branching ratio,
since there is no known cascade process for $\psi^\prime$
production.
By using again  $\Gamma_T/\Gamma =  0.3$,
we estimate
$
\xi_{\psi^\prime} \simeq 0.18.
$
Experimental data
are not available yet for
the other
 $\psi$ resonances, $\rho$ and
$\omega$.
Encouraged by $\xi_\psi$ and $\xi_{\psi^\prime}$
above, we
 assume  the same $a_3$, $b_3$ and take $\xi_V = \xi_\psi=
\sqrt{|a_3|^2 + |b_3|^2} \simeq 0.2 $ for all
vector mesons $V=\psi, \, .... \rho,  \; \omega$.
Then the VMD amplitude for the charmonium states is
\begin{equation}
A_T^\psi( b \rightarrow d\, \gamma) =
 e \frac{G_F}{\sqrt{2}}
v_c e_\psi\,
  \sum_i \left(
\frac{f^2_{\psi_i}(0)}{m_b}
\right) \left[
a_3 \, \bar d \sigma^{\mu\nu} (1-\gamma_5)\, q_\nu  b
+ b_3\, \bar d  \sigma^{\mu\nu} (1+\gamma_5) \, q_\nu
b \right] \, \epsilon_\mu^\gamma.
\label{LD:amplitude-psi}
\end{equation}
Note that now the decay constants are extended to
$q^2 \rightarrow 0$. That gives rise to
a suppression factor $f^2_{\psi_i}(0)= k\,
f^2_{\psi_i}(m_{\psi_i}^2)$
where we take $k \simeq
0.12$ from Ref.~\cite{Deshpande:LDPenguins,terasaki};
no suppression factor is taken for
$\rho$ and $\omega$, whose masses
are smaller.
By substituting in~(\ref{LD:amplitude-psi})
$v_c
\rightarrow v_u$,
$e_\psi \rightarrow   e_{\rho(\omega)} 1/\sqrt{2} $,
$ f^2_\psi({m^2}_\psi)~\rightarrow~f^2_{\rho(\omega)}(m_{\rho}^2) $,
$k \rightarrow 1$ we obtain
$ A_T^{\rho(\omega)}( b \rightarrow d\, \gamma). $

The contributions
to the amplitude
coming from $\psi$ and its resonances
are approximately equal
to the contributions coming from $\rho $ and
$\omega$ intermediate states times the CKM factors
\begin{equation}
\frac{|A_T^{\rho}+A_T^{\omega}|}{|A_T^\psi|} \simeq
1.06 \; \frac{ |v_u |}{
 | v_c |} \simeq 0.4.
\end{equation}

The sign between the $\psi$ mode and the $\omega+\rho$ mode
may not be reliable, since, at least
in the exclusive case  $\psi\,K^\ast$, we expect
sizable  final state interactions.
However, if we ignore possible final state phases,
the contribution
proportional to $v_u$ coming from
charmonium states (by unitarity
$v_c=-v_t-v_u$)
 approximately cancel
 the contribution coming from
$\rho$ and $\omega$~\cite{Deshpande:LDPenguins}.

The short distance amplitude at the leading order in QCD
corrections is
\begin{equation}
A^{SD}(b \rightarrow d \gamma) = e\, \frac{G_F}{\sqrt{2}}
v_t  C_7^{eff}(m_b)
\frac{1}{4\pi^2} m_b {\bar d} \sigma^{\mu\nu} (1+\gamma_5) \,q_\nu
           b \, \epsilon_\mu^\gamma,
\label{LD:amplitude-sd}
\end{equation}
where $C_7^{eff}(m_b) \simeq 0.3$
from Eqs (\ref{sd-coefficients}) and (\ref{vaffa}).

The decay rate from
(\ref{LD:amplitude-psi})
alone is about
$10^{-3}$ of the short
distance decay rate,
in agreement with \cite{Deshpande:LDPenguins}.
The change to decay rate from short
distance due to long distance contributions proportional
to $v_u$ is less than $ 6
 \% $
with respect to the short distance
decay rate,
while the change
due to long distance contributions proportional
to $v_u$ is less than $ 12
 \%  \times |v_u|/|v_t|$.
The change in the decay rate due to $A_T^{\rho}+{A_T}^\omega$
is less than $ 6
 \% \times |v_u|/|v_t| $.
Therefore,
the ratio between the short and the long distance pieces is  small
with respect to the theoretical errors
in  the short distance
decay rate itself
which has  been estimated to be about
$25 \%$ ~\cite{buras:theor-uncertainties}.

\section{Acknowledgments}

It is a pleasure to thank
Ahmed Ali, Howard Georgi, Mitch Golden, Amarjit Soni,
Raman Sundrum, Hitoshi Yamamoto
for interesting discussions.
This work has been supported by a INFN fellowship and partially
by a NSF grant PHY-92-18167.


\newpage
\begin{table}
\begin{center}
\begin{tabular}{ c c c } \hline \hline
  $V$ &
  $\Gamma_{V \rightarrow e^+\,e^-}\, (\gev) $ \cite{particle-data-book} &
  $f_V \, (\gev) $
\\ \hline
$ \rho^0(0.770) $ & $ 6.74 \times 10^{-6} $ & 0.216   \\ \hline
$ \omega(0.782) $ & $ 0.60 \times 10^{-6} $ & 0.194   \\ \hline
$ \psi(3.097)   $ & $ 5.27 \times 10^{-6} $ & 0.406   \\ \hline
$ \psi(3.685)   $ & $ 2.44 \times 10^{-6} $ & 0.301   \\ \hline
$ \psi(3.770)   $ & $ 2.83 \times 10^{-7} $ & 0.104   \\ \hline
$ \psi(4.040)   $ & $ 7.28 \times 10^{-7} $ & 0.172   \\ \hline
$ \psi(4.160)   $ & $ 0.77 \times 10^{-6} $ & 0.180   \\ \hline
$ \psi(4.415)   $ & $ 0.47 \times 10^{-6} $ & 0.145   \\ \hline \hline
\label{tableld}
\end{tabular}
\caption{Values of the measured decay rate
 of the vector meson $V$ into $e^+\, e^-$ pair
and of the decay constant $f_V$.
The $V$ decay constant is defined by
$ <0|J_\mu^{em}|V>
\equiv e_V\, m_V \, f_V(m^2_V) \epsilon_\mu^V $
and is calculated
from the experimental data for
$\Gamma(V\rightarrow e^+\,e^-)$,
according to the formula
$
\Gamma(V \rightarrow  e^+\,e^-) =\frac{4 \pi \alpha^2}{3}
\frac{e_V^2\,f_V^2(m_V^2)}{m_V}\;\left[1-4\,\frac{m_e^2}{m_V^2}\right]^{1/2}
\, \left[1+2\, \frac{m_e^2}{m_V^2}\right].
$}
\end{center}
\end{table}

\end{document}